\documentclass[aps,prd,onecolumn,amsmath,amssymb,nofootinbib,preprintnumbers,12pt,letterpaper]{revtex4}
\pdfoutput=1

\usepackage{amsmath,latexsym,amssymb,amsfonts}
\usepackage[pdftex]{color,graphicx}
\usepackage{bm}
\usepackage{gensymb}
\newcommand{\blue}[1]{\textcolor{blue}{#1}}

\begin{document}

\rightline{YITP-22-47}

\title{Resolving information loss paradox with Euclidean path integral$^*$}%\footnote{Honorable Mention in Gravity Research Foundation 2022 Awards for Essays on Gravitation.}

\author{
\textsc{Pisin Chen$^{1,2}$}\footnote{{\tt pisinchen{}@{}phys.ntu.edu.tw; Corresponding author}},
\textsc{Misao Sasaki$^{1,4,5}$}\footnote{{\tt misao.sasaki{}@{}ipmu.jp}},
\textsc{Dong-han Yeom$^{6,7}$}\footnote{{\tt innocent.yeom{}@{}gmail.com}}
and
\textsc{Junggi Yoon$^{8,9,10}$}\footnote{{\tt junggiyoon{@}gmail.com}}
}

\affiliation{
$^{1}$\footnotesize{Leung Center for Cosmology and Particle Astrophysics, National Taiwan University, Taipei 10617, Taiwan}\\
$^{2}$\footnotesize{Department of Physics and Graduate Institute of Astrophysics, National Taiwan University, Taipei 10617, Taiwan}\\
%$^{c}$\footnotesize{Graduate Institute of Astrophysics, National Taiwan University, Taipei 10617, Taiwan}\\
%$^{3}$\footnotesize{Kavli Institute for Particle Astrophysics and Cosmology,
%SLAC National Accelerator Laboratory, Stanford University, Stanford, California 94305, USA}\\
$^{4}$\footnotesize{Kavli Institute for the Physics and Mathematics of the Universe (WPI), University of Tokyo, Chiba 277-8583, Japan}\\
$^{5}$\footnotesize{Yukawa Institute for Theoretical Physics, Kyoto University, Kyoto 606-8502, Japan}\\
$^{6}$\footnotesize{Department of Physics Education, Pusan National University, Busan 46241, Republic of Korea}\\
$^{7}$\footnotesize{Research Center for Dielectric and Advanced Matter Physics, Pusan National University, Busan 46241, Republic of Korea}\\
$^{8}$\footnotesize{Asia Pacific Center for Theoretical Physics, Pohang 37673, Republic of Korea}\\
$^{9}$\footnotesize{Department of Physics, POSTECH, Pohang 37673, Republic of Korea}\\
$^{10}$\footnotesize{School of Physics, Korea Institute for Advanced Study, Seoul 02455, Republic of Korea}
}

\begin{abstract}
\centering\begin{minipage}{\dimexpr\paperwidth-5cm}
\vspace{4mm}
The information loss paradox remains unresolved ever since Hawking's seminal discovery of black hole evaporation. 
In this essay, we revisit the entanglement entropy via
Euclidean path integral (EPI) and allow for the branching of semi-classical
histories during the Lorentzian evolution. We posit that there exist two histories that contribute 
to EPI, where one is information-losing that dominates at early times, while the other is information-preserving that dominates at late times. By
so doing we recover the Page curve and preserve the unitarity, albeit with the Page time shifted significantly
towards the late time. One implication is that the entropy bound may thus be violated. We compare 
our approach with string-based islands and replica wormholes concepts.
\end{minipage}
\end{abstract}

\maketitle

%\vskip 2.0in
%\vspace{-2mm}
\centerline{$^*$Honorable Mention in Gravity Research Foundation 2022 Awards for Essays on Gravitation.}

%\newpage

%\tableofcontents

\newpage

\subsection*{Introduction}

The information loss paradox \cite{Hawking:1976ra} remains an unresolved problem in modern theoretical physics since Hawking's seminal discovery of black hole evaporation in 1975 \cite{Hawking:1974sw}. This problem reveals a potential inconsistency between unitary quantum mechanics and general relativity.

According to Hawking's computation \cite{Hawking:1974sw}, a black hole will emit radiation and evaporate completely, 
where the radiation depends only on mass, charge, and angular momentum of the black hole. This implies the \textit{loss of information}. 
If this indeed happens, then one would lose the predictable power of quantum mechanics. 
On the other hand, if one believes that unitarity should not be violated, then the information loss paradox must be resolved and explained. 

%Now one can ask, \textit{why is this paradox so difficult to resolve?} 
To resolve this paradox, one typically invokes one or more of the following widely accepted assertions  \cite{Yeom:2009zp}: 
(1) \textit{unitarity} of the black hole evaporation process, 
(2) \textit{general relativity}, at least far from the singularity, 
(3) \textit{local quantum field theory} in a semi-classical black hole background spacetime, 
and (4) \textit{equivalence} of the Bekenstein-Hawking entropy and the Boltzmann entropy.
%and (5) \textit{existence of an observer} who can perform several thought experiments.
There have been arguments, however, that not all of these assertions are consistent \cite{Yeom:2009zp,Almheiri:2012rt}. In particular, a soft spot appears to be the notion of \textit{entanglement entropy} and the \textit{monogamic nature} of the maximal entanglement \cite{Maldacena:2013xja}. Evidently, entanglement entropy, and thus Assertion~(4), is the prime suspect. 

In this essay, we propose a new resolution to the information loss problem a la the notion of \textit{wave function of the universe} cast in the Euclidean path integral formulation. By insisting the multi-history condition and the late-time dominance condition, which will be explained later, we manage to recover the Page curve, albeit being necessarily modified with the Page time shifted significantly towards the late-time of black hole evaporation. Surprisingly, our resolution is a natural outcome of the most conservative approach to quantum gravity, i.e., canonical quantum gravity \cite{DeWitt:1967yk}, without resorting to exotic means.

\subsection*{Entanglement entropy and Page time}

To quantitatively describe the flow of information, the introduction of the \textit{entanglement entropy} is found very useful \cite{Page:1993wv}. 
Let us consider a system composed of two subsystems $A$ and $B$, and a pure state given by $| \Psi \rangle$. 
The density matrix of the system is $\rho \equiv | \Psi \rangle \langle \Psi |$. The reduced density matrix for the subsystem $A$ is given by
tracing the subsystem $B$ out, i.e., $\rho_{A} \equiv \mathrm{tr}_{B} \rho$.
Likewise, the reduced density matrix for the subsystem $B$ is given by
$\rho_{B} \equiv \mathrm{tr}_{A} \rho$. 
Then the von Neumann entropy of the subsystem $A$ is $S_{B}(A) \equiv - \mathrm{tr}_{A} \rho_{A} \ln \rho_{A}$. 
This is known as the entanglement entropy of a subsystem $A$,  which is the same as that of its complementary, $S_A(B)$, if the state is pure.

\begin{figure}
\begin{center}
\includegraphics[scale=0.6]{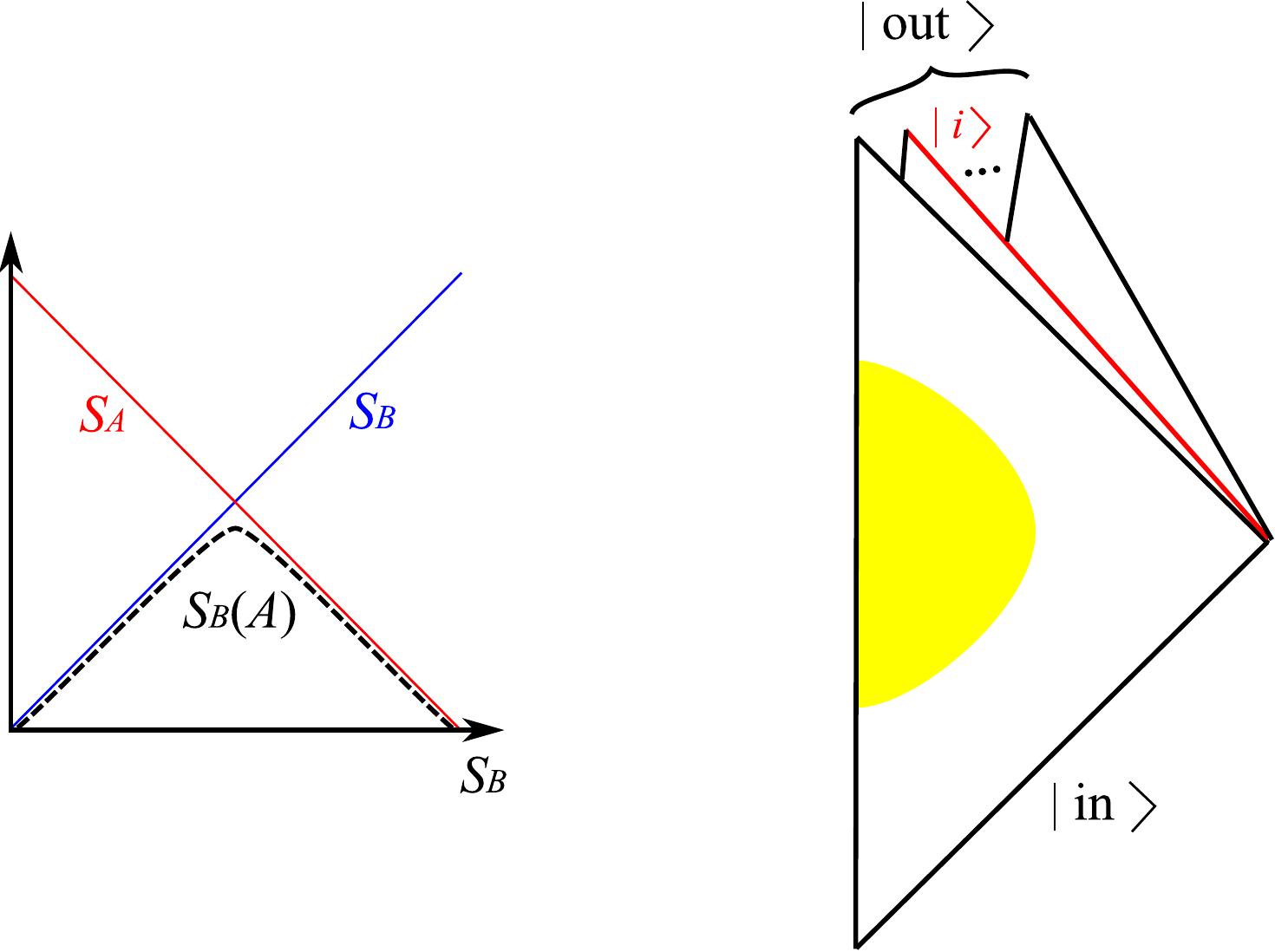}
\caption{\label{fig:Page}Left: The Page curve, where $S_{A}$ and $S_{B}$ are Boltzmann entropies of $A$ and $B$, respectively, and $S_{B}(A)$ denotes the entanglement entropy. Right: The path integral from the in-state $| \mathrm{in} \rangle$ to the out-state $| \mathrm{out} \rangle$, where the out-state is a superposition of classical boundaries $\{ | i \rangle \}$.}
\end{center}
\end{figure}

Let us consider quantum states of a black hole. We divide the system into the interior of a black hole, denoted by $A$, and the Hawking radiation in the exterior, denoted by $B$.
We assume that initially all degrees of freedom were in $A$, and, as time goes on, they are monotonically transmitted from $A$ to $B$ by the Hawking radiation. 
According to the analysis by Page \cite{Page:1993df}, by assuming a typical pure state with a fixed number of total degrees of freedom in the beginning, 
the entanglement entropy is almost the same as the Boltzmann entropy of the radiated particles. 
However, when the original entropy of the black hole decreased to approximately its half value, the entanglement entropy of the radiation begins to decrease (left of Fig.~\ref{fig:Page}).
This turning time is called the \textit{Page time}. If one further assumes that the Boltzmann entropy of the black hole is the same as the Bekenstein-Hawking entropy, one can compute the value of the Page time, which is approximately $\sim M^{3}$ (in Planck units), 
where $M$ is the black hole mass; it is evident that even at this time, the black hole remains semi-classical.

\subsection*{Trouble with entanglements}

Based on general relativity (2) and semi-classical quantum field theory (3), the Hawking radiation can be interpreted as a particle-antiparticle pair 
creation process, where the particle and antiparticle are maximally entangled \cite{Hwang:2017yxp}. On the other hand, for a pure and random system with fixed degrees of freedom (1), 
after the initial Bekenstein-Hawking entropy decreased to its half-value (4), 
the entanglement entropy should begin to decrease. 
In other words, the Hawking particles
after the Page time starts to be entangled, in addition, with the previously emitted Hawking particles.

Let us now assume that there is an observer who detects a Hawking particle, say $\alpha$, after the Page time. 
The observer will notice that the particle $\alpha$ is entangled with the previously emitted Hawking particle $\beta$, following the entanglement distillation protocol.
On the other hand, this $\alpha$ must be maximally entangled with its antiparticle partner, say $\alpha'$, if the infalling observer sees a smooth horizon. 
This is contradictory because $\alpha$, which is maximally entangled with $\alpha'$, cannot be entangled with $\beta$ 
due to the monogamy nature of maximal entanglements (1) \cite{Maldacena:2013xja}. How can this inconsistency be resolved?

\subsection*{Wave function of the universe and superposition of states}

According to canonical quantum gravity \cite{DeWitt:1967yk}, the entire information of the universe is included by the wave function of 
the universe $\Psi$, which is a functional of the three-geometry $h_{\mu\nu}$ and a matter field configuration, say $\phi$, 
on top of $h_{\mu\nu}$. 
This wave function should satisfy the quantum Hamiltonian constraint equation, the so-called Wheeler-DeWitt (WDW) equation. 
Since this is the fundamental equation of quantum gravity, unitarity must be manifest.

One can assume that the in-state of the wave function of the universe is given by $| \mathrm{in} \rangle \equiv | h_{\mu\nu}^{(\mathrm{in})}, \phi^{(\mathrm{in})} \rangle$, where we assume that this in-state is a fixed classical configuration. This configuration will evolve to an out-state, say $| \mathrm{out} \rangle \equiv | h_{\mu\nu}^{(\mathrm{out})}, \phi^{(\mathrm{out})} \rangle$. The WDW equation will determine the wave function of the universe for a given initial condition.
%$\langle \mathrm{out} | \mathrm{in} \rangle = \langle h_{\mu\nu}^{(\mathrm{out})}, \phi^{(\mathrm{out})} | h_{\mu\nu}^{(\mathrm{in})}, \phi^{(\mathrm{in})} \rangle$.

In order to understand the black hole evaporation process, we need to choose a proper out-state. 
It should be such that the observer at future infinity will see a semi-classical spacetime. 
However, the final out-state is not necessarily a unique classical spacetime; rather, 
it can be a \textit{superposition of states corresponding to each classical spacetime} \cite{Hartle:2015bna}. This follows from the observation that 
the Hawking radiation can be interpreted as a result of quantum tunneling \cite{Chen:2018aij}. 
Thus
\begin{eqnarray}
\left|\mathrm{out} \right\rangle = \sum_{i,\alpha} c_{\alpha,i} \left|\alpha;i \right\rangle,
\end{eqnarray}
where $|\alpha;i \rangle$ is a quantum state associated with a semi-classical state labeled by $i$\blue{,} while
 $\alpha$ represents microscopic quantum degrees of freedom in the semi-classical state,
and $c_{\alpha,i} = \langle \alpha;i|\mathrm{in}\rangle$ (right of Fig.~\ref{fig:Page}).

\begin{figure}
\begin{center}
\includegraphics[scale=0.7]{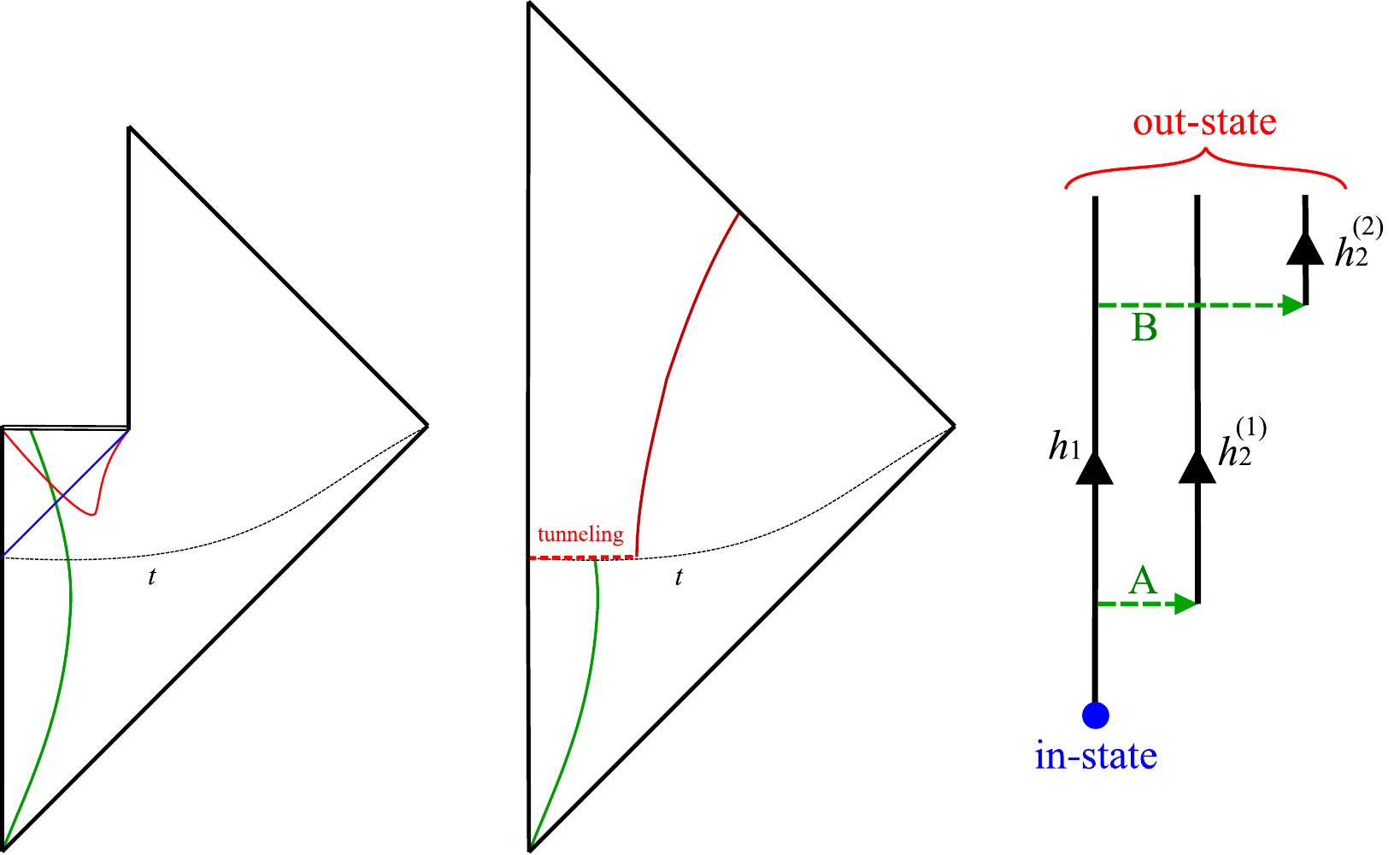}
\caption{\label{fig:Page2}Left: the causal structure of the usual semi-classical black hole, where the green curve is the trajectory of the collapsing matter, the red curve is the apparent horizon, and the blue line is the event horizon. Middle: the causal structure after a quantum tunneling at the time slice $t$. After the tunneling, matter or information (red curve) is emitted and the black hole structure disappears. Right: $h_{1}$ is the information-losing history, while $h_{2}^{(1,2)}$ are the information-preserving histories. Tunneling may happen either early time (A) or late time (B); the tunneling probability must be dominated at the late time.}
\end{center}
\end{figure}

\subsection*{Essential conditions for unitary entanglement entropy}

As mentioned in the above, a natural consequence from the picture based on canonical quantum gravity is 
that the out-state is a superposition of semi-classical states. 
Let us assume that one can categorize the semi-classical states into two distinct classes.
The first class is those with \textit{information-losing histories}, where the black hole keeps existing
and loses information by the Hawking radiation, and therefore the entanglement entropy monotonically 
increases up to the end point (left of Fig.~\ref{fig:Page2}). 
The second class is those with \textit{information-preserving histories}, 
which appear as a result of quantum tunneling, where there is no black hole, hence no singularity nor event horizon.
Hence the entanglement entropy is zero for this class of histories (middle of Fig.~\ref{fig:Page2}).

When the in-state is dominated by information-losing histories, the (semi-classical) observer outside of the black hole cannot have access to the degrees of freedom inside of the black hole, which leads to the increase of the entanglement entropy. After the Page time, if the out-state is dominated by information-preserving histories, the observer can now have access to all degrees of freedom which would not have been measured in the information-losing histories. Hence, the entanglement entropy will vanish in the end. 
To summarize, in order to obtain a unitary Page curve, what one needs to justify 
is the following two essential conditions (right of Fig.~\ref{fig:Page2}) \cite{Chen:2021jzx}: 
\begin{itemize}
\item[--] 1. \textit{Multi-history condition}: existence of multiple information-preserving and non-preserving histories;
\item[--] 2. \textit{Late-time dominance condition}: dominance of the information-preserving history at late-time.
\end{itemize}
For simplicity and without losing generality, let us assume that there are only two histories, one information-losing 
and the other information-preserving. 
Then one can approximately evaluate the entanglement entropy as
$S_{\mathrm{ent}} \simeq p_{1} S_{1} + p_{2} S_{2}$, where $1$ denotes the information-losing history,
 $2$ denotes the information preserving-history, 
$p_{i}$ and $S_{i}$ ($i=1,\,2$) are the probability and the entanglement entropy for each history, 
respectively. 
In the beginning, the history $1$ dominates, and hence explains the increasing phase of
 the entanglement entropy. 
However, at late times, the history $2$ dominates as $S_{2}=0$.
 The total entanglement entropy will eventually decrease to zero.

\subsection*{Realization: Euclidean path integral approach}

To confirm these essential conditions, one needs to compute the transition element
 $\langle i | \mathrm{in} \rangle$, where $|i \rangle$ is a state representing a classical history.
Here we have omitted the label $\alpha$ for the microscopic degrees of freedom 
in each classical history for notational simplicity. 
The probability of each history is given by $p_{i} = |\langle i | \mathrm{in} \rangle|^{2}$.
If we recover the label $\alpha$, then we have $p_i=\sum_\alpha|\langle\alpha; i | \mathrm{in} \rangle|^{2}$.

Although we do not yet have a final formulation for quantum gravity, at the semi-classical level
the Euclidean path integral can provide a good approximation that captures the essence of a bona fide full-blown quantum gravity theory \cite{Hartle:1983ai}:
\begin{eqnarray}
\langle i | \mathrm{in} \rangle = \langle h_{\mu\nu}^{(i)}, \phi^{(i)} | h_{\mu\nu}^{(\mathrm{in})}, \phi^{(\mathrm{in})} \rangle = \int \mathcal{D} g_{\mu\nu}\mathcal{D}\Phi \; e^{-S_{\mathrm{E}}[g_{\mu\nu}, \Phi]},
\end{eqnarray}
where $S_{\mathrm{E}}$ is the Euclidean action and all Euclidean geometries that connect
$| h_{\mu\nu}^{(\mathrm{in})}, \phi^{(\mathrm{in})} \rangle$ to $| h_{\mu\nu}^{(i)}, \phi^{(i)} \rangle$ are summed over. 
This path integral can be well approximated by summing over on-shell solutions, 
i.e., either Lorentzian classical solutions or Euclidean instantons.

To evaluate the evolution of the entanglement entropy, the following two
technical observations are important:
\begin{itemize}
\item[--] To interpret the Hawking radiation as a quantum tunneling process, 
one can consider the perturbation of a free scalar field in the 
Euclidean Schwarzschild background~\cite{Chen:2018aij}. 
The tunneling probability is given by $p \simeq e^{-2B}$, 
where $B = S_{\mathrm{E}} (\mathrm{solution}) - S_{\mathrm{E}} (\mathrm{background})$.
If we impose the reality condition on the scalar field at future infinity, 
the solution becomes complex-valued. 
Nevertheless, there is some evidence that we may accept such a complex instanton 
as a legitimate classical solution that dominates the path integral \cite{Hartle:2007gi}.

Assuming that the emitted energy $\omega$ is negligible compared to the black hole mass $M$, 
the probability is approximately found as $e^{-\omega/T}$ with the Hawking temperature
 $T = 8\pi M$ \cite{Chen:2018aij}.
 This shows that the Hawking radiation may indeed be regarded as a quantum tunneling process.
If we extrapolate this result to the case when the whole mass is emitted, 
then the spacetime transits to Minkowski spacetime, with its probability given by $e^{-S}$ where 
$S = 4\pi M^{2}$. Namely, there exists a tunneling channel to a trivial geometry.
In fact, one can construct such an instanton by considering
 a thin-wall scalar field configuration~\cite{Chen:2017suz}.

\item[--] Adopting the above thin-shell toy model as the tunneling channel to a trivial geometry,
one realizes that there are instantons with multiple periods in the Euclidean time 
as long as one can correctly perform analytic continuation to the future infinity. 
Hence, these multiple period solutions should be taken into account
for the computation of the tunneling probability from the information-losing history to
the information-preserving history. 
One obtains the following:
\begin{eqnarray}
2B_n &=& n \left( (\mathrm{bulk\; term\; of\; solution}) + (\mathrm{boundary\; term\; of\; solution}) \right) \nonumber \\
 &&- (\mathrm{boundary\; term\; of\; background}),
\end{eqnarray}
where $n=1,\, 2,\,3\,\cdots$.  
\end{itemize}

\begin{figure}
\begin{center}
\includegraphics[scale=0.6]{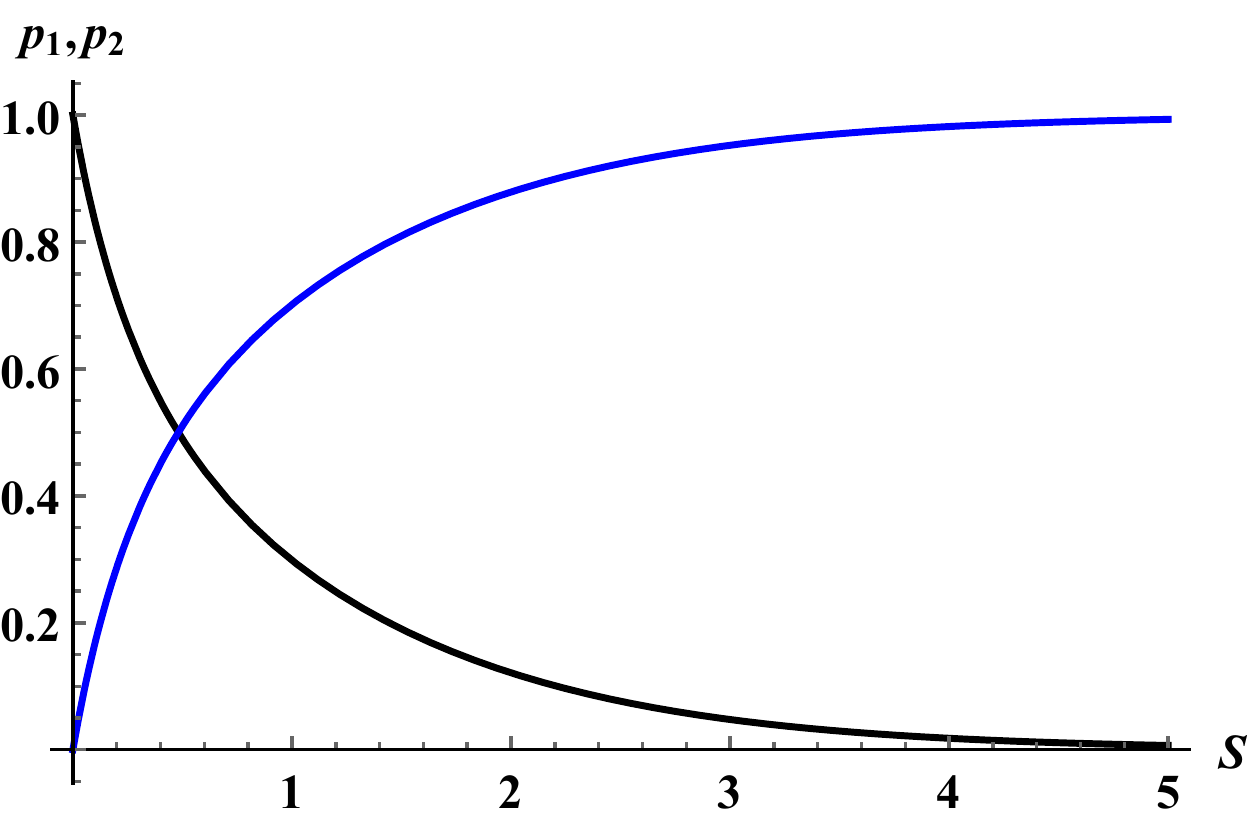}
\includegraphics[scale=0.8]{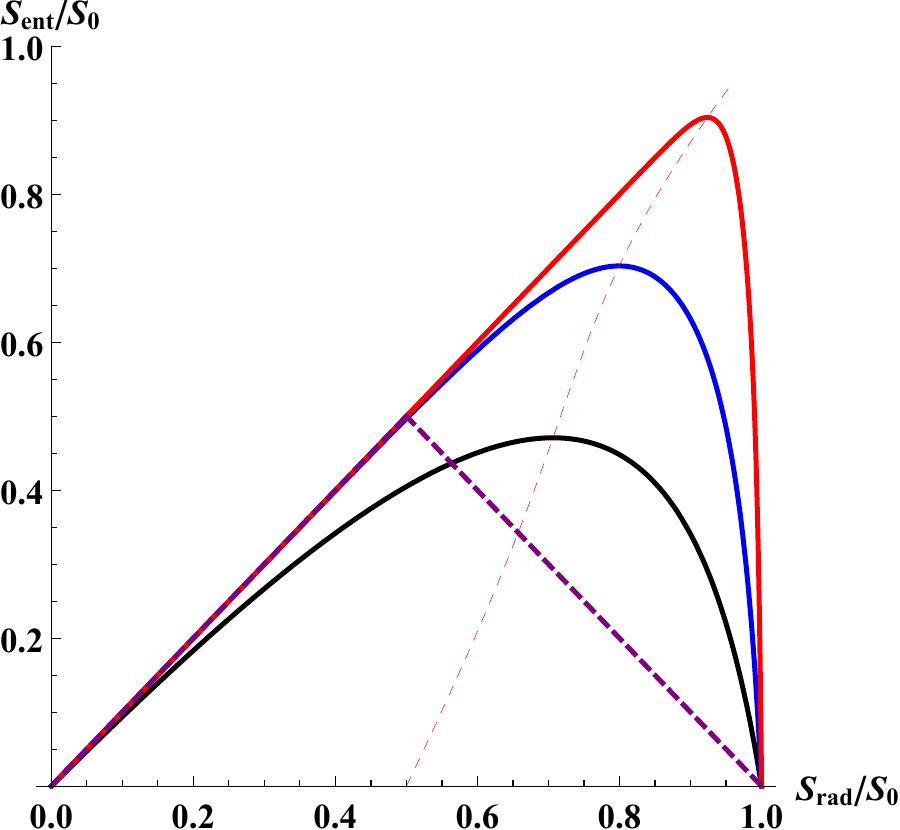}
\caption{\label{fig:prob}Left: Probabilities of the information-losing history ($p_{1}$, blue curve) and the information-preserving history ($p_{2}$, black curve). Right: Entropy of emitted radiation $S_{\mathrm{rad}}/S_{0}$ vs. entanglement entropy $S_{\mathrm{ent}}/S_{0}$, where $S_{0} = 3$ (black), $10$ (blue), $50$ (red), respectively. The thin red dashed curve is the location of the Page time, i.e., $dS_{\mathrm{ent}}/dS_{\mathrm{rad}} = 0$.}
\end{center}
\end{figure}

It is now straightforward to evaluate the entanglement entropy. 
The tunneling probability is given by \cite{Chen:2021jzx}
\begin{eqnarray}
\frac{p_{2}}{p_{1}} = \sum_{n=1}^{\infty} e^{-(2n-1)S} = \frac{1}{e^{S} - e^{-S}}.
\end{eqnarray}
Interestingly, although the $n = 1$ solution makes the most dominant contribution,
the multi-period instanton contributions become important as $M$ decreases.
If we impose the normalization condition $p_{1} + p_{2} = 1$, we obtain
\begin{eqnarray}
p_{1} = \frac{e^{S} - e^{-S}}{1 + e^{S} - e^{-S}}, \quad p_{2} = \frac{1}{1 + e^{S} - e^{-S}}.
\end{eqnarray}
This successfully explains that $p_{1}$ initially dominates and $p_{2}$ is exponentially 
suppressed, while at the late times,  $p_{2}$ eventually dominates over $p_{1}$ as $S$ decreases. 
This explains the late-time dominance condition (left of Fig.~\ref{fig:prob}).

Thus if we assume that the entanglement entropy of the history 1 (information-losing history) 
monotonically increases and the entanglement entropy of the history 2 (information-preserving history) 
is zero, we obtain the expectation value of the entanglement entropy, or that 
from the point of view of the entire wave function, as
\begin{eqnarray}
S_{\mathrm{ent}} = p_{1} \times (S_{0} - S) + p_{2} \times 0
 = (S_{0} - S) \left( \frac{e^{S} - e^{-S}}{1 + e^{S} - e^{-S}} \right),
\end{eqnarray}
where $S_{0}$ is the initial Boltzmann entropy of the black hole (right of Fig.~\ref{fig:prob}).

Now we can explain the unitary evolution of the black hole system:
The entanglement entropy starts from zero, monotonically increases for a while, 
and eventually decreases to zero. 
An important, interesting observation is that the turning point is far beyond the moment 
when the Bekenstein-Hawking entropy decreased to its half value. It occurs when
$S\sim \log S_{0}$.
In other words, the equivalence of the Bekenstein-Hawking entropy and the Boltzmann entropy is violated.
However, one has to realize that this equivalence is not a result of any fundamental principles. 
In fact, an intriguing counter example has recently been pointed out
in \cite{Buoninfante:2021ijy}, which is perfectly consistent with both local field theory and 
semi-classical general relativity. 
In our case, it is also important to note that the turning point occurs while the
semi-classical approximation is still perfectly valid provided that the initial
black hole is macroscopic, $S\sim \log S_{0}\gg 1$. (For example, even for a black hole of
a fairly small mass $M\sim10^5$g, $S_0\sim 10^{20}$ and hence $S\sim 46$.)

\subsection*{Recent developments based on string theory}

It is interesting to compare our approach with the recent developments based on string theory, where the entanglement entropy was investigated by evaluating several quantum extremal surfaces (QES). It was found~\cite{Almheiri:2019psf,Almheiri:2019hni}, in two-dimensional gravity, that a saddle of QES without an \emph{island} is dominant before the Page time, which induces the increase of the entanglement entropy. After the Page time, QES is dominated by the other saddle with an island, which results in the decrease of the entanglement entropy. It was argued in~\cite{Almheiri:2020cfm} that such a role played by the island can also reproduce the Page curve in higher dimensional black holes.

The string-based islands approach and ours share two essential features. One, there exist more than two contributions to the evolution of the entanglement entropy, where one is dominant at early-time and the other at late-time. Two, the final entanglement entropy is dominated by the late-time condition. That is, the multi-history condition and the late-time dominance condition in our Euclidean path integral approach are analogous to the spirit of the QES computations for black hole.

However, the physical interpretation of the islands conjecture is still not very clear. One reason is that the entanglement entropy computed in QES is based on the density matrix, instead of the quantum states, which is the standard quantum field theory approach and what we have followed. It is therefore natural to ask, what is the implication of the islands or the replica wormholes \cite{Almheiri:2019qdq,Penington:2019kki} to the more orthodox, state-level path integral in Lorentzian signatures, and vice versa?

%At this moment, there is no clear dependence between the instantons that we mentioned and replica wormholes. However, one may optimistically expect that the Euclidean path integral approach may shed some lights to the correct interpretation; the replica wormholes may correspond to a kind of trivial geometry that explains a Lorentzian spacetime without singular nor event horizon, or at least, explains a violation of the classical equations of motion.

\subsection*{Conclusion and future prospects}

We have argued that the canonical quantum gravity with the Euclidean path integral approach can provide a consistent picture to resolve the information loss paradox. By computing the wave function of the universe with the Euclidean path integral, we successfully justified the two essential conditions, that is, the \textit{multi-history condition} and the \textit{late-time dominance condition}, and eventually obtained a modified Page curve that preserves the unitarity but with the Page time shifted significantly towards the late-time. 

%Then, where is information, and how can we recognize it? Perhaps, a primitive answer is that information is already outside of the black hole from the beginning, but one cannot yet see the entire history at the beginning; as such semi-classical observers like us will \textit{effectively} experience the loss of information \cite{Hawking:2005kf}.

Note that the entanglement entropy of a black hole can never exceed its Boltzmann entropy. Therefore, if one insists on Assertion (4) (equivalence between the Bekenstein-Hawking entropy and the Boltzmann entropy), then the entanglement entropy cannot exceed the Bekenstein-Hawking entropy. In contrast, one salient outcome of our computation is that there exists a moment where the entanglement entropy is greater than the Bekenstein-Hawking entropy. Assertion (4) is thus violated in our approach. This necessarily implies that the number of states inside the horizon must have been accumulated during the black hole evaporation, although such an accumulation is strictly bounded. We emphasize that this violation of Assertion (4) is not in contradiction with basic principles of physics \cite{Buoninfante:2021ijy}. nor the first three assertions.

In our picture, the turning point of the Page curve, though shifted significantly towards the end-life of the black hole evaporation, is still in the semi-classical regime of quantum gravity as we have shown. Hence there might be a way to experimentally investigate our notion. If our model can be examined not only by theoretical means, but also by experimental methods \cite{Chen:2015bcg}, then the synergy between theory and experiment may hopefully lead us to the ultimate understanding of the information loss paradox.

%\newpage

\section*{Acknowledgments} PC is supported by Taiwan's Ministry of Science and Technology (MOST) and the Leung Center for Cosmology and Particle Astrophysics (LeCosPA), National Taiwan University. MS is supported in part by JSPS KAKENHI grant Nos. 19H01895, 20H04727, and 20H05853. DY is supported by the National Research Foundation of Korea (Grant no.: 2021R1C1C1008622, 2021R1A4A5031460). JY is supported by the National Research Foundation of Korea (Grant no.: 2019R1F1A1045971, 2022R1A2C1003182). JY is also supported by an appointment to the JRG Program at APCTP through the Science and Technology Promotion Fund, the Lottery Fund of the Korean government, and by Gyeongsangbuk-do and Pohang-si.


\newpage

\begin{thebibliography}{200}

%\cite{Hawking:1976ra}
  \bibitem{Hawking:1976ra}
  S.~W.~Hawking,
  %``Breakdown of Predictability in Gravitational Collapse,''
  Phys. Rev. D \textbf{14}, 2460-2473 (1976).

%\cite{Hawking:1974sw}
  \bibitem{Hawking:1974sw} 
  S.~W.~Hawking,
  %``Particle Creation by Black Holes,''
  Commun.\ Math.\ Phys.\  {\bf 43}, 199 (1975)
  [Commun.\ Math.\ Phys.\  {\bf 46}, 206 (1976)].
  %doi:10.1007/BF02345020, 10.1007/BF01608497
  %%CITATION = doi:10.1007/BF02345020, 10.1007/BF01608497;%%

%\cite{Yeom:2009zp}
  \bibitem{Yeom:2009zp}
  D.~Yeom and H.~Zoe,
  %``Semi-classical black holes with large N re-scaling and information loss
  %problem,''
  Int. J. Mod. Phys. A \textbf{26} (2011), 3287-3314
  %doi:10.1142/S0217751X11053924
  [arXiv:0907.0677 [hep-th]].

%\cite{Almheiri:2012rt}
  \bibitem{Almheiri:2012rt}
  A.~Almheiri, D.~Marolf, J.~Polchinski and J.~Sully,
  %``Black Holes: Complementarity or Firewalls?,''
  JHEP \textbf{02} (2013), 062
  %doi:10.1007/JHEP02(2013)062
  [arXiv:1207.3123 [hep-th]].

%\cite{Maldacena:2013xja}
  \bibitem{Maldacena:2013xja}
  J.~Maldacena and L.~Susskind,
  %``Cool horizons for entangled black holes,''
  Fortsch. Phys. \textbf{61}, 781-811 (2013)
  %doi:10.1002/prop.201300020
  [arXiv:1306.0533 [hep-th]].

%\cite{DeWitt:1967yk}
\bibitem{DeWitt:1967yk}
  B.~S.~DeWitt,
  %``Quantum Theory of Gravity. 1. The Canonical Theory,''
  Phys. Rev. \textbf{160} (1967), 1113-1148.
  %doi:10.1103/PhysRev.160.1113


%\cite{Page:1993wv}
  \bibitem{Page:1993wv}
  D.~N.~Page,
  %``Information in black hole radiation,''
  Phys. Rev. Lett. \textbf{71} (1993), 3743-3746
  %doi:10.1103/PhysRevLett.71.3743
  [arXiv:hep-th/9306083 [hep-th]].

%\cite{Page:1993df}
  \bibitem{Page:1993df}
  D.~N.~Page,
  %``Average entropy of a subsystem,''
  Phys. Rev. Lett. \textbf{71}, 1291-1294 (1993)
  %doi:10.1103/PhysRevLett.71.1291
  [arXiv:gr-qc/9305007 [gr-qc]].

%\cite{Hwang:2017yxp}
\bibitem{Hwang:2017yxp}
  J.~Hwang, H.~Park, D.~Yeom and H.~Zoe,
  %``How can we erase states inside a black hole?,''
  J. Korean Phys. Soc. \textbf{73} (2018) no.10, 1420-1430
  %doi:10.3938/jkps.73.1420
  [arXiv:1712.00347 [hep-th]].

%\cite{Hartle:2015bna}
  \bibitem{Hartle:2015bna}
  J.~Hartle and T.~Hertog,
  %``Quantum transitions between classical histories,''
  Phys. Rev. D \textbf{92}, no.6, 063509 (2015)
  %doi:10.1103/PhysRevD.92.063509
  [arXiv:1502.06770 [hep-th]].

%\cite{Chen:2018aij}
  \bibitem{Chen:2018aij} 
  P.~Chen, M.~Sasaki and D.~Yeom,
  %``Hawking radiation as instantons,''
  Eur.\ Phys.\ J.\ C {\bf 79}, no. 7, 627 (2019)
  %doi:10.1140/epjc/s10052-019-7138-0
  [arXiv:1806.03766 [hep-th]].
  %%CITATION = doi:10.1140/epjc/s10052-019-7138-0;%%

%\cite{Chen:2021jzx}
  \bibitem{Chen:2021jzx}
  P.~Chen, M.~Sasaki, D.~Yeom and J.~Yoon,
  %``Solving information loss paradox via Euclidean path integral,''
  [arXiv:2111.01005 [hep-th]].

%\cite{Hartle:1983ai}
  \bibitem{Hartle:1983ai} 
  J.~B.~Hartle and S.~W.~Hawking,
  %``Wave Function of the Universe,''
  Phys.\ Rev.\ D {\bf 28}, 2960 (1983)
  [Adv.\ Ser.\ Astrophys.\ Cosmol.\  {\bf 3}, 174 (1987)].
  %doi:10.1103/PhysRevD.28.2960
  %%CITATION = doi:10.1103/PhysRevD.28.2960;%%

%\cite{Hartle:2007gi}
  \bibitem{Hartle:2007gi}
  J.~B.~Hartle, S.~W.~Hawking and T.~Hertog,
  %``No-Boundary Measure of the Universe,''
  Phys. Rev. Lett. \textbf{100} (2008), 201301
  %doi:10.1103/PhysRevLett.100.201301
  [arXiv:0711.4630 [hep-th]].

%\cite{Chen:2017suz}
  \bibitem{Chen:2017suz} 
  P.~Chen, G.~Dom\`enech, M.~Sasaki and D.~Yeom,
  %``Thermal activation of thin-shells in anti-de Sitter black hole spacetime,''
  JHEP {\bf 1707}, 134 (2017)
  %doi:10.1007/JHEP07(2017)134
  [arXiv:1704.04020 [gr-qc]].
  %%CITATION = doi:10.1007/JHEP07(2017)134;%%

%\cite{Buoninfante:2021ijy}
  \bibitem{Buoninfante:2021ijy}
  L.~Buoninfante, F.~Di Filippo and S.~Mukohyama,
  %``On the assumptions leading to the information loss paradox,''
  JHEP \textbf{10}, 081 (2021)
  %doi:10.1007/JHEP10(2021)081
  [arXiv:2107.05662 [hep-th]].

%\cite{Almheiri:2019psf}
  \bibitem{Almheiri:2019psf}
  A.~Almheiri, N.~Engelhardt, D.~Marolf and H.~Maxfield,
  %``The entropy of bulk quantum fields and the entanglement wedge of an   evaporating black hole,''
  JHEP \textbf{12}, 063 (2019)
  %doi:10.1007/JHEP12(2019)063
  [arXiv:1905.08762 [hep-th]].

  %\cite{Almheiri:2019hni}
\bibitem{Almheiri:2019hni}
A.~Almheiri, R.~Mahajan, J.~Maldacena and Y.~Zhao,
%``The Page curve of Hawking radiation from semiclassical geometry,''
JHEP \textbf{03}, 149 (2020)
%doi:10.1007/JHEP03(2020)149
[arXiv:1908.10996 [hep-th]].
%253 citations counted in INSPIRE as of 16 Sep 2021

%\cite{Almheiri:2020cfm}
\bibitem{Almheiri:2020cfm}
A.~Almheiri, T.~Hartman, J.~Maldacena, E.~Shaghoulian and A.~Tajdini,
%``The entropy of Hawking radiation,''
Rev. Mod. Phys. \textbf{93}, 035002 (2021)
%doi:10.1103/RevModPhys.93.035002
[arXiv:2006.06872 [hep-th]].

%\cite{Penington:2019kki}
\bibitem{Penington:2019kki}
G.~Penington, S.~H.~Shenker, D.~Stanford and Z.~Yang,
%``Replica wormholes and the black hole interior,''
[arXiv:1911.11977 [hep-th]].
%306 citations counted in INSPIRE as of 16 Sep 2021

%\cite{Almheiri:2019qdq}
  \bibitem{Almheiri:2019qdq}
  A.~Almheiri, T.~Hartman, J.~Maldacena, E.~Shaghoulian and A.~Tajdini,
  %``Replica Wormholes and the Entropy of Hawking Radiation,''
  JHEP \textbf{05}, 013 (2020)
  %doi:10.1007/JHEP05(2020)013
  [arXiv:1911.12333 [hep-th]].

%\cite{Chen:2015bcg}
  \bibitem{Chen:2015bcg}
  P.~Chen and G.~Mourou,
  %``Accelerating Plasma Mirrors to Investigate Black Hole Information Loss Paradox,''
  Phys. Rev. Lett. \textbf{118}, no.4, 045001 (2017)
  %doi:10.1103/PhysRevLett.118.045001
  [arXiv:1512.04064 [gr-qc]].

\end{thebibliography}
\end{document}